# Antiadiabatic View of Fast Environmental Effects on Optical Spectra

D. K. Andrea Phan Huu, Rama Dhali, Carlotta Pieroni, Francesco Di Maiolo, Cristina Sissa, Francesca Terenziani, and Anna Painelli[*]

*Department of Chemistry, Life Sciences and Environmental Sustainability, Parma University, Parma 43124, Italy*

An antiadiabatic approach is proposed to model how the refractive index of the surrounding medium affects optical spectra of molecular systems in condensed phases. The approach solves some of the issues affecting current implementations of continuum solvation models and more generally of effective models where a classical description is adopted for the molecular environment.

The definition of reliable and practical models to simulate how the local environment affects the spectra of molecular materials represents a theoretical and computational challenge with enormous practical implications: environmental effects can be detrimental to the performance of molecular materials for advanced applications, including organic light emitting device and solar cells, but, when properly understood, they can be exploited towards optimized materials in a smart-matrix approach.

If specific interactions can be neglected, electrostatic forces dominate the interplay between the solute (the molecule of interest) and the solvent (the local molecular environment, either a solvent, a solid matrix, a biological environment, etc.), so that the solvent is described in terms of its dielectric properties [1]. A good solvent is transparent in the spectral window of interest (typically the visible and near-UV, ~1–4 eV) and its electronic absorption bands are found deep in the UV region (> 6 eV). Kramers-Krönig equations relate the real and imaginary parts of the dielectric response: since in the region of interest the solvent is transparent, in the same region the dielectric constant is real and almost frequency independent, its square root being usually referred to as the refractive index, $\epsilon_{\text{opt}} = \eta^2$ [2]. Vibrational transitions are weak and marginally contribute to the dielectric constant. Orientational motions are optically silent in nonpolar solvents, whose dielectric constant therefore stays almost invariant down to its static value: $\epsilon_{\text{st}} \sim \epsilon_{\text{opt}}$. In polar solvents instead the orientational motion of solvent molecules gives a large contribution to the static dielectric constant and $\epsilon_{\text{st}} > \epsilon_{\text{opt}}$.

With respect to relevant solute degrees of freedom (d.o.f.), typically in the visible and near UV spectral regions, the electronic d.o.f. of the solvent (with resonances deep in the UV) are much faster, while the orientational motion of polar solvent molecules (picosecond timescale in liquid solvents, longer in solid matrices) is much slower. In either case, the distinctly different timescales for the solute and solvent dynamics allow for the separation of the two systems in effective solvation models. Two families of models have been developed in this context: in the QM-MM approach, the solute is modeled by a quantum-mechanical (QM) Hamiltonian accounting for the electrostatic potential generated by the surrounding medium that is described in a classical molecular mechanics (MM) approach. In a simpler but powerful approach, the solvent is described as a continuum dielectric medium, as in the polarizable continuum model (PCM) or in the conductor-like screening model (COSMO) [3–9].

In current implementations of effective solvation models both fast and slow environmental d.o.f. are treated in the adiabatic approximation, neglecting the corresponding kinetic energies and solving the solute problem for fixed configuration of the surrounding medium. This approximation works very well for polar solvation, associated with d.o.f. much slower than the solute d.o.f., but it fails when applied to the fast electronic d.o.f. of the solvent, whose kinetic energy cannot be disregarded. Fast d.o.f. should rather be treated in an *antiadiabatic* (AA) approach [10], describing the limit where the electronic clouds of solvent molecules instantaneously respond to the charge fluctuations in the solute [11–13]. Here a model for fast solvation is introduced that, amenable to a numerically exact solution, is used to validate the AA approach and to critically review current implementations of continuum solvation models.

We describe the solute-solvent interaction in the dipolar approximation: the solvent generates at the solute location an electric field (the *reaction field*) which, in turn, is proportional to the solute dipole moment. This self-consistent model set the basis to understand solvatochromism [14,15] was adopted in parametric models [13,16–18], and was used to discuss PCM implementations [3,4]. At the equilibrium, both the fast and slow components of the reaction field are proportional to the expectation value of the solute dipole moment in the state of interest, $(\vec{F}_{el/or})_{eq} = r_{el/or} \langle \hat{\vec{\mu}} \rangle$, with the proportionality constant determined by the medium dielectric properties [15,19]:

$$r_{el} = \frac{2}{4\pi\epsilon_0 a^3} f(\epsilon_{opt}), \qquad r_{or} = \frac{2}{4\pi\epsilon_0 a^3} [f(\epsilon_{st}) - f(\epsilon_{opt})] \quad (1)$$

where $a$ is the radius of the (spherical) cavity occupied by the solute, $\epsilon_0$ is the vacuum permittivity, and $f(\epsilon) = (\epsilon - 1)/(2\epsilon + 1)$. Modeling the solvent as an elastic medium and enforcing the equilibrium condition, the Hamiltonian of the solvated molecule reads [13,20]:

$$H = H_{gas} + \left[\frac{\vec{F}_{el}^2}{2r_{el}} + T_{el} - \hat{\vec{\mu}}\vec{F}_{el}\right] + \left[\frac{\vec{F}_{or}^2}{2r_{or}} - \hat{\vec{\mu}}\vec{F}_{or}\right] \quad (2)$$

where $H_{gas}$ is the gas phase molecular Hamiltonian and the two parentheses group terms relevant to the electronic and orientational solvation. $T_{el}$ is the kinetic energy associated with the electronic reaction field. The corresponding term in the second parenthesis is missing since the adiabatic approximation works well for the orientational field. In the following we only address electronic solvation, shortly discussing polar solvation in the concluding section. Moreover, we consider quasilinear molecules, whose dipole moment has sizable matrix elements only along a special molecular axis (Fig. 1), at least for the states of interest. $F_{el}$ and $\hat{\mu}$ denote the main components of the reaction field and of the dipole moment operator, respectively.

In second-quantization we set $F_{el} = g(\hat{b}^\dagger + \hat{b})$, where $\hat{b}$ ($\hat{b}^\dagger$) is the boson annihilation (creation) operator, $g = \sqrt{\hbar\omega_{el}r_{el}/2}$ and $\omega_{el}$ is the frequency associated with the solvent electronic polarization (typically in the ultraviolet). With these definitions, the Hamiltonian of a molecule only coupled to $F_{el}$ [the first two terms in the Hamiltonian in Eq. (2)] reads:

$$H_0 = H_{gas} - g\hat{\mu}(\hat{b}^\dagger + \hat{b}) + \hbar\omega_{el}\left(\hat{b}^\dagger\hat{b} + \frac{1}{2}\right) \quad (3)$$

If $H_{gas}$ is defined on a finite basis set $(|f_1\rangle, |f_2\rangle, ..., |f_N\rangle)$, a numerically exact nonadiabatic solution of the Hamiltonian in Eq. (3) is obtained diagonalizing the Hamiltonian matrix written on the direct product basis: $(|f_1\rangle, |f_2\rangle, ..., |f_N\rangle) \times (|0\rangle, |1\rangle, |2\rangle, ...)$, where $|n\rangle$ are the eigenstates of the harmonic oscillator in the last term of Eq. (3) [20]. Of course, the infinite oscillator basis is truncated to large enough $n$ as not to affect the properties of interest.

In the proposed model, a single effective oscillator with frequency $\omega_{el}$ describes the electronic spectrum of the solvent [20]. This oversimplified view only applies if the solvent absorption bands occur at much higher energy than the solute absorption bands, so that the details of the solvent spectrum become irrelevant. This is not a specific limitation of our model, but it is an intrinsic limitation of any effective solvation model. Indeed, if the details of the electronic excitation spectrum of the solvent are important, then a complete QM description of the solute and the surrounding medium is unavoidable. Consistently with the separation of the solute-solvent dynamics, we then adopt an AA approach (Fig. 2) [11–13], setting $\omega_{el} \to \infty$ to obtain the following Hamiltonian (details of the derivation in the Supplemental Material [20]):

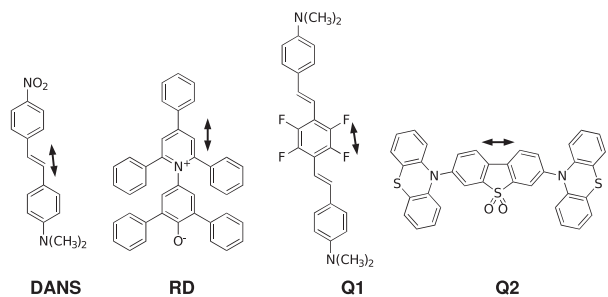

FIG. 1. The molecules considered in this Letter. The arrows mark the direction of the main component of the dipole moment operator [20]. DANS (dimethylamino-nitrostylbene) and RD (the Reichardt dye) are polar dyes; Q1 (a fluorinated bis-alkylaminostyryl derivative) and Q2 [3,7-bis(10H-phenothiazin-10-yl)dibenzo[b,d]thiophene-S,S-dioxide] are representative quadrupolar dyes.

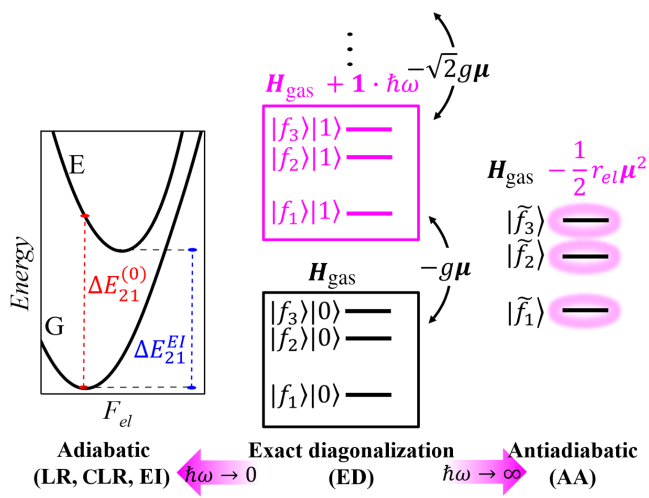

FIG. 2. Central panel: a sketch of the nonadiabatic Hamiltonian. Left panel: in the adiabatic approximation we show the ground and excited state potential energy curves and the definition of the vertical and EI transition energies. Right panel: a sketch of the AA renormalized molecular states and Hamiltonian.

$$H_{AA} = H_{gas} - \frac{r_{el}}{2}\hat{\mu}^2. \quad (4)$$

This equation applies to quasilinear molecules, its analogous for 3D structures has $\hat{\mu}^2$ substituted by $\hat{\vec{\mu}}^2$.

To validate our view, following a similar strategy as in Ref. [21], a few state model (FSM) is defined for the four molecules in Fig. 1. DANS and RD are polar (electron-donor-acceptor, DA) dyes, showing positive and negative solvatochromism, respectively [22,23]. Q1 and Q2 are quadrupolar (DAD) dyes: both have negligible polarity but Q1, of interest for nonlinear optics [18], has a sizable transition dipole moment to the first excited state, while Q2, of interest for thermally activated delayed fluorescence [24], has a negligible transition dipole moment. We run gas phase TD-DFT [CAM-B3LYP, 6-31G(d)] calculations and select the first three singlets as the molecular basis $|f_1\rangle, |f_2\rangle, |f_3\rangle$ [20]. The matrix elements of the dipole moment operator are calculated by MULTIWFN software [25]. Figure 3 compares the molecular properties calculated in the AA approximation and upon exact diagonalization, ED, of $H_0$ [Eq. (3)] setting $\hbar\omega_{el} = 6$ or 20 eV. Results are plotted against $f(\epsilon_{opt})$, estimated for each molecules setting $a$ to the relevant Onsager radius [20,26].

Results in Fig. 3 confirm that the AA Hamiltonian in Eq. (4) represents the $\omega_{el} \to \infty$ limit of the nonadiabatic Hamiltonian in Eq. (3). Moreover, with the notable exception of Q1, results are marginally affected by the specific $\omega_{el}$ value, suggesting that effective solvation models are reliable even for solvents with comparatively low-energy excitations (for most organic media 6 eV represents the absorption cutoff, but the absorption maxima are located at much higher energies [20]). For Q1, a highly polarizable dye, the solute-solvent separation is more delicate and should be considered with care in largely polarizable environments.

The ground state dipole moment of the two polar dyes, DANS and RD, smoothly increases with $f(\epsilon_{opt})$, due to the stabilization of polar states in condensed media. For DANS, a polar dye with a neutral ground state, this implies an increase of the transition dipole moment and a decrease of the transition frequency [16,22], while the opposite occurs for RD, a dye with a zwitterionic ground state [16]. Quadrupolar dyes, Q1 and Q2, have vanishing permanent dipole moment, but the solvent polarizability is responsible for a sizable decrease of the transition frequency.

Current implementations of effective solvation models adopt the adiabatic approximation to deal with fast solvation. For comparison purposes, we solve the Hamiltonian in Eq. (3) in the adiabatic approximation, adopting the same strategies as implemented in GAUSSIAN16 [26]. The first step is the calculation of the ground state obtained upon diagonalization of the adiabatic Hamiltonian with $F_{el}$ fixed at the ground state equilibrium. The top panels of Fig. 4 compare the adiabatic and AA estimates of DANS and RD permanent dipole moments (Q1 and Q2 have vanishing dipole moments). The adiabatic approximation fails already in the calculation of the ground state. In particular, the adiabatic approximation underestimates the increase of the ground state dipole moment of DANS in condensed media. Indeed, the ground state dipole moment of DANS is smaller than its excited state dipole moment [22]. The reaction field equilibrated at the ground state is therefore small and more polar states than the ground state are less stabilized in the adiabatic approximation than in the AA approach where each state is stabilized by the interaction with its own reaction field. The opposite occurs for RD, whose dipole moment is larger in the ground than in the excited state [23].

Turning attention to spectral properties, in the linear response (LR) approach the transition energy is calculated from the vertical transition energy, $\Delta E_{21}^{(0)}$ (see Fig. 2) as follows [3,4]:

$$\Delta E_{21}^{LR} = \Delta E_{21}^{(0)} - r_{el}|\mu_{21}|^2 \quad (5)$$

The LR transition energy in Fig. 4 compares well with the AA result only for DANS. In general the LR energies are not accurate since they do not account for the variation of the solute polarity upon excitation [3–5]. To improve on LR, state specific approaches were introduced [5,6,9]. Among them, the external iteration (EI) approach equilibrates the fast solvation field around the excited state and calculates the transition energy as the energy difference between the equilibrated excited and ground states (Fig. 2) [5]. Calculated EI energies always deviate considerably from AA results. More fundamentally, EI suffers from a basic flaw when applied to fast solvation, since the optimized ground and excited states are eigenstates of two different adiabatic Hamiltonians, thus precluding the calculation of transition dipole moments. The corrected linear response (CLR) approach circumvents this problem only accounting for perturbative corrections to transition energies, while maintaining the wave funtions unperturbed. In CLR the correction to the transition energy is proportional to the square of the variation of the dipole moment upon excitation [6,9,20]:

$$\Delta E_{21}^{\text{CLR}} = \Delta E_{21}^{(0)} - \frac{r_{\text{el}}}{2}(\mu_{22} - \mu_{11})^2. \quad (6)$$

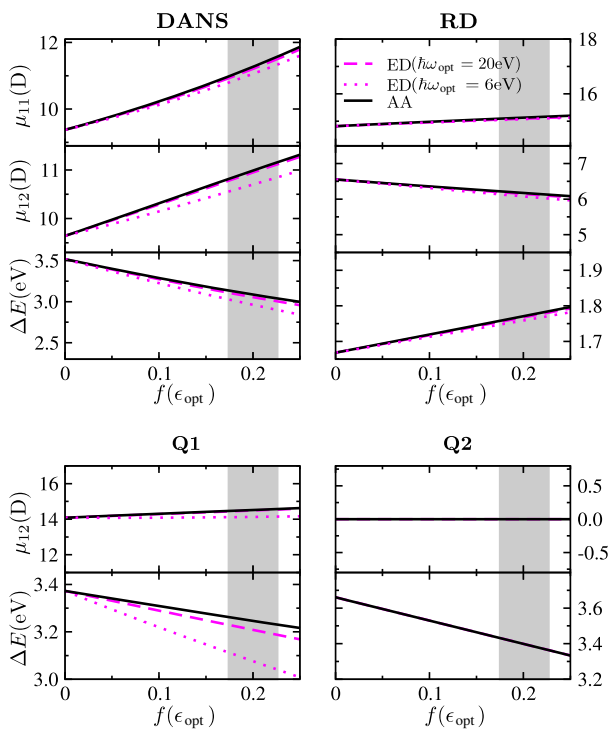

FIG. 3. Top panels: for the two polar dyes the ground state dipole moment $\mu_{11}$, the transition dipole moment $\mu_{12}$ and the transition energy $\Delta E$ are reported vs $f(\epsilon_{opt})$. Bottom panels: for quadrupolar dyes the transition dipole moment $\mu_{12}$ and the transition energy $\Delta E$ are reported vs $f(\epsilon_{opt})$. Black lines refer to AA results, magenta lines show ED results, obtained for $\omega_{el} = 6$ and 20 eV (dotted and dashed lines, respectively). For Q2 all lines are superimposed. The shaded area marks the region where most organic solvents are located.

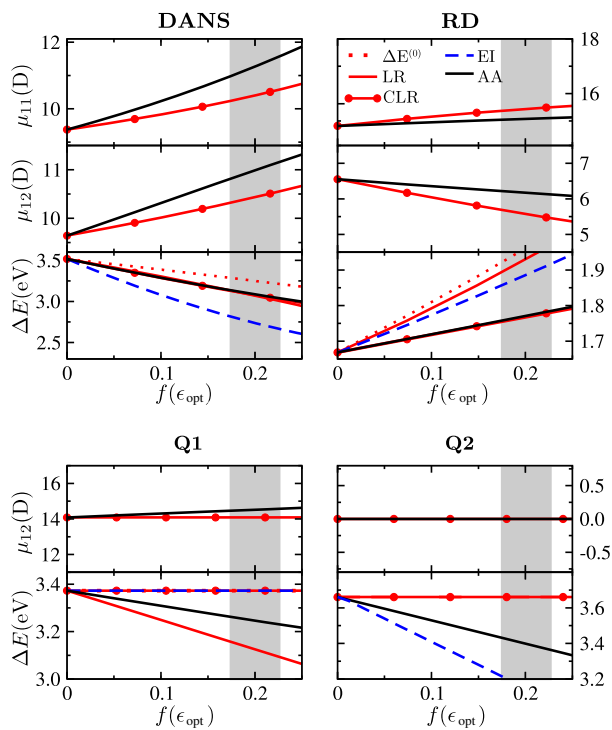

FIG. 4. The same as in Fig. 3 but comparing AA results (black line) with adiabatic results (colored lines). The ground state dipole moment $\mu_{11}$ (red line with dots) is the same in all adiabatic implementations. The transition dipole moment $\mu_{12}$ is undefined in EI, and is the same for LR and CLR approaches. For transition energies, the dotted red lines show the vertical excitation energy, the continuous red lines show LR results, the continuous red lines with dots show CLR results, the blue line show the EI results. For DANS, LR and CLR energies are almost superimposed. For Q1 the vertical excitation energy, CLR and EI energies are coincident. For Q2 all adiabatic energies but EI are superimposed. The shaded area marks the region relevant for organic solvents and matrices.

The CLR estimate of the transition energies is good for the two polar dyes, whose solvatochromic shifts are governed by the variation of the molecular dipole moment upon excitation [14]. Some error cancellation on the two dipole moments clearly enters into play here, since, as discussed above, the adiabatic estimate of the ground state dipole moment is poor.

Adiabatic approaches fail in the most striking way for the quadrupolar dyes, Q1 and Q2. These dyes have a negligible polarity and therefore have vanishing CLR corrections. The sizable transition dipole moment of Q1 leads to a sizable LR correction, indeed largely deviating from AA results. Q2 instead has a negligible transition dipole moment, then for this dye both LR and CLR corrections vanish. Neither LR nor CLR reproduce the excited state stabilization of Q2 due to the medium refractive index. The solvent polarizability indeed stabilizes instantaneous charge fluctuations in the solute, an effect that cannot be appreciated in any adiabatic approach to fast solvation.

To validate the proposed FSM, in the Supplemental Material [20] adiabatic results in Fig. 4 are compared with analogous results from TD-DFT calculations for solvated dyes adopting the adiabatic implementations of PCM in GAUSSIAN16 [26]. The comparison confirms that the adopted FSM captures the basic physics of our systems. The only interesting observation is that sizable CLR corrections to the transition energies of the two quadrupolar dyes are calculated in TD-DFT. Since Q1 and Q2 are nonpolar, these corrections are due to quadrupolar and higher order terms in the solute-solvent interactions, that are fully disregarded in our model. However, the important point here is not the quality of the dipolar approximation. Indeed our results demonstrate that the adiabatic approximation fails in the most dramatic way to describe fast solvation since it cannot account for the first order (dipolar) corrections to the transition energy of nonpolar dyes.

The limits of current implementations of continuum solvation models are known [4,6,9], here we demonstrate that they are rooted in the adiabatic treatment of fast solvation. Adopting different approximation schemes (LR, CLR, EI, etc.) for the calculation of transition energies cannot cure the basic problem: the adiabatic approximation does not account for the fast fluctuations of the solvent electronic clouds in response to the charge fluctuations in the solute and therefore cannot provide a reliable description of the effects of the medium refractive index on molecular properties and spectra. This problem, addressed here with specific reference to continuum solvation models, affects more generally all effective solvation models where the solvent is described classically, including the QM-MM approach. In QM-MM, even when accounting for the polarizability of the medium, a state-specific adiabatic Hamiltonian is defined for the solute and is diagonalized in the presence of a frozen potential due to the surrounding medium.

Effective solvation models must rely on an AA description of environmental electronic d.o.f., leading to a renormalized AA molecular Hamiltonian that accounts for the instantaneous response of the solvent electronic clouds to charge fluctuations in the solute. The eigenstates of the AA Hamiltonian directly enter the calculation of optical spectra, without the need to invoke state-specific Hamiltonians, quite naturally solving the conundrum of calculating transitions between states obtained upon diagonalization of different Hamiltonians.

Once fast solvation is accounted for in the AA Hamiltonian, polar solvation can be dealt with in the adiabatic approximation. For this application EI [5], leading to formally exact results, is more accurate than either LR or CLR approaches, based on perturbative expansions [6,9]. Optical transitions occur vertically with respect to slow d.o.f. [27]. Accordingly, the eigenstates involved in the absorption process are obtained diagonalizing the adiabatic Hamiltonian with the potential due to slow solvation fixed to the ground state equilibrium value. Similarly, the states involved in fluorescence are obtained diagonalizing the adiabatic Hamiltonian with the slow-solvation potential equilibrated to the lowest excited singlet. In either case, transitions are calculated between states that are obtained from the diagonalization of the same EI Hamiltonian. The issue of incongruent eigenstates, affecting EI when applied to fast solvation, does not show up in dealing with polar solvation, for which the adiabatic approximation works well.

Extending the model to multipolar terms in the solute-solvent interaction is certainly feasible, but we believe that, having properly framed the problem of fast solvation, reliable AA effective Hamiltonians will be developed towards realistic and detailed descriptions of the molecular systems. The $GW$-Bethe-Salpeter equation formalism coupled to continuum solvation models [28–30] is promising in this respect, but the development of reliable approaches to fast solvation to be implemented into popular TD-DFT computational codes is highly desirable.

This project received funding from the European Union Horizon 2020 research and innovation programme under Grant Agreement No. 812872 (TADFlife), from the Italian Ministero dell'Istruzione, dell'Università e della Ricerca (MIUR) through the Grant "Dipartimenti di Eccellenza" (DM 11/05/2017 n. 262). This work was supported by CINECA through projects IscrC_iiCT-MMM and IscrC_CANTA and benefits from the HPC (High Performance Computing) facility of the University of Parma, Italy.